\documentclass[12pt,preprint]{aastex}

\shorttitle{}
\shortauthors{}

\begin{document}

\title{Extreme X-ray spectral variability in the Seyfert 2 Galaxy NGC 1365 
} 

\author{G. Risaliti\altaffilmark{1,2}, M. Elvis\altaffilmark{1}, 
G. Fabbiano\altaffilmark{1},
A. Baldi\altaffilmark{1}, A. Zezas\altaffilmark{1}}
\email{grisaliti@cfa.harvard.edu}

\altaffiltext{1}{Harvard-Smithsonian Center for Astrophysics, 60 Garden St. 
Cambridge, MA 02138 USA}
\altaffiltext{2}{INAF - Osservatorio di Arcetri, L.go E. Fermi 5,
Firenze, Italy}
\begin{abstract}
We present multiple {\em Chandra} and {\em XMM-Newton} observations of
the type 1.8 Seyfert Galaxy NGC 1365, which shows the most dramatic X-ray spectral changes 
observed so far in an AGN: the source switched from 
reflection dominated to transmission dominated and back in just 6 weeks. 
During this time the soft thermal component, arising from a $\sim1$~kpc region
around the center, remained constant. The reflection component is constant at all
timescales, and its high flux relative to the primary component
implies the presence of thick gas covering a large
fraction of the solid angle. 
The presence of this gas, and the fast variability time scale, suggest that
the Compton-thick to Compton thin change is due to variation in the line-of-sight
absorber, rather than to extreme intrinsic emission variability. We discuss
a structure of the circumnuclear absorber/reflector which can explain
the observed X-ray spectral and temporal properties.

  \end{abstract}

\keywords{ Galaxies: AGN --- X-rays: galaxies --- Galaxies: individual (NGC 1365)}

\section{Introduction}

According to the Unified Model of Active Galactic Nuclei (AGNs, see
review by Urry \& Padovani, 1995), an axisymmetric
absorber/reflector is present around the central black hole of AGNs,
with a size between that of the Broad Emission Line Region (BELR,
$\sim 1000~R_S$, Schwarzschild radii, corresponding to $\sim10^{-2}$~pc
for a $10^8~M_\odot$ black hole) and that of the Narrow Emission Line region
(tens to hundreds parsecs).  The simplest geometrical and physical
configuration of such absorber is that of a homogeneous torus (Krolik \& Begelman 1988)
beyond the dust sublimation radius ($r_{sub}\sim1.3L_{UV,46}^{-0.5}$~pc, 
Barvanis 1987, where
$L_{UV,46}$ is the UV luminosity in units of $10^{46}$~erg~s$^{-1}$).
However, this view has been recently challenged by several pieces of
observational evidence:
\\ 
(1) Dramatic X-ray absorbing column density changes (factors of $>$10)
over a few years have been seen in several type~2 (narrow permitted
line) Seyferts (Risaliti, Elvis \& Nicastro 2002), 
ruling out a homogeneous absorber.  \\
(2) Rapid column density variability, on time scales of hours,
requires an X-ray absorber no larger than the BELR. Such changes have
been detected in the brightest absorbed Seyfert Galaxy, NGC~4151
($\sim10-30$~hours, Puccetti et al. 2004) and in the Seyfert~2 Galaxy NGC~4388
(4~hours, Elvis et al. 2004). Assuming that the absorber is made of
material moving with Keplerian velocity around the central source,
these observations imply that its distance is of the order of that of
the broad line clouds, i.e. $\sim10^3~R_S$, where $R_S=2GM_{BH}/c^2$ is
the Schwartzschild radius.  \\ 
(3) The reflection components in the X-ray spectra of Compton thin
Seyfert Galaxies are systematically stronger than expected from
reflection off gas with the same column density measured in
absorption. This has been shown convincingly for a few sources in
which a detailed measurement of the reflection component has been
possible (NGC~1365, Risaliti et al. 2000; NGC~2992, Gilli et
al. 2002; NGC~6300, Guainazzi et al. 2003) and in a statistical sense
in a sample of $\sim 20$ bright Seyfert 2s (Risaliti 2002).

NGC~1365 ($z$=0.0055) 
is a particularly striking example of extreme X-ray variability and
highly efficient reflection. It was observed by ASCA in
1995 (Iyomoto et al. 1997) 
in a reflection-dominated state (Thomson optical depth $\tau>1$,
corresponding to N$_H>1.5\times10^{24}$~cm$^{-2}$) 
and, three years later, in a Compton-thin state
($N_H\sim4\times10^{23}$~cm$^{-2}$) by BeppoSAX (Risaliti et
al. 2000).  The 2-10 keV flux of the reflection
emission was more than 5\% that of the intrinsic
spectrum measured by BeppoSAX. 
This corresponds to a ratio R=1 between the normalizations of the reflected 
and direct components in the {\tt PEXRAV} reflection model (Magdziarz \& Zdziarski 1995)
Such a high reflection efficiency can be achieved only if a thick
(N$_H>3\times 10^{24}$~cm$^{-2}$) reflector covers a large fraction of
the solid angle around the central source (Ghisellini, Haardt \& Matt
1994). 

Similar switching from reflection-dominated 
to transmission dominated states 
has now been observed in a handful sources on timescales of a few years
(Matt, Guainazzi \& Maiolino 2004)
and could be due either to
extreme variations of the intrinsic luminosity and so of the ionization state
of the absorber (as suggested by Matt et al. 2004),
or to Compton-thick clouds moving across the line of sight, so changing the line of sight
column density\footnote{It is worth noticing the conceptual difference between
``reflection dominated'' (which merely refers to the observed spectral shape) and
``Compton thick'' (which implies a physical interpretation of the observed
reflection dominated spectrum as due to Compton thick absorption of the 
direct component).}.

Here we present an X-ray observational campaign on
NGC 1365, consisting of 3 {\em XMM-Newton} observations, plus
a {\em Chandra} observation (Obsid 3554) performed just three weeks before
the first {\em XMM-Newton} observation.
The observing times
are relatively short ($\sim10-20$~ksec), 
therefore a temporal analysis within single observations is not
possible. We will therefore discuss the spectra extracted from the
whole observations. 
Two more, longer (60~ksec), {\em XMM-Newton} observations were recently 
performed, as a continuation of our program of monitoring of this source.
Detailed spectral analysis of all the {\em XMM-Newton} observations 
will be presented in a forthcoming
paper (Risaliti et al. 2005, in prep.)

Here we will concentrate on the analysis of the most striking
result of our study, i.e. dramatic hard X-ray continuum variations:
changes from
reflection-dominated to transmission-dominated 
spectra occurred on times scale
shorter than three weeks.
During this time, the soft thermal component remained constant in all
observations.  The high resolution of {\em Chandra} allowed us to
resolve this emitting region, which extends over $\sim 1$~kpc from the
center, while the hard component originates in a region $<200$~pc
(2~arcsec) diameter\footnote{Throughout this paper we adopt
$H_0=70$~km~s$^{-1}$~Mpc$^{-1}$ (Spergel et al.~2003).}.\\

The observation log of the X-ray observations of NGC~1365 is
reported in Table~1 together with basic spectral fitting results (\S 2).

\section{Data Analysis and Results}

The spectra presented here have been obtained with CCD detectors:
the ACIS-S instrument
on board {\em Chandra}, and the EPIC
PN and MOS instruments on board {\em XMM-Newton}.
All observations were performed in full-frame mode. 
We did not find any significant pile-up. This was expected for
{\em XMM-Newton} (the previous published 2-10~keV flux is $<10^{11}$~erg~s$^{-1}$~cm$^{-2}$,
mostly emitted at energies $>3$~keV) , while for {\em Chandra} this is a consequence of the
extremely faint state of the source (Table~1) at that time.

The data were reduced using the standard procedures, using the CIAO v.3.0
%
%
package for the {\em Chandra} observation and SAS v6.0
%
%
%
 for the {\em
XMM-Newton} observations.  The spectra were extracted from  circular
regions of a radius of 2" ({\em Chandra}) and 30" (both {\em Chandra}
and {\em XMM-Newton}). 
The background spectra were extracted from regions close
to our source and free from contamination by bright serendipitous sources.
The background contribution is negligible for the {\em Chandra} spectrum of
the central 2"~radius region (thanks to the small extraction region), and 
for the two {\em XMM-Newton} spectra when the source was
caught in a bright state (XMM~1 and XMM~3). 
In the remaining two spectra (XMM~2, and the {\em Chandra} spectrum from the
30"~radius region)
the background contamination is $\sim10$\%.

The spectra were analyzed using the XSPEC 12.0
%
%
%
package. The spectra 
of the {\em XMM-Newton}  observations are
shown in Fig.~1. 
A visual inspection of this figure 
is sufficient to clearly notice the main spectral variation:
in the 2-10~keV band the XMM~2 spectrum is an order of magnitude weaker than the
XMM~1 and XMM~3 observations and has a prominent iron line
with an equivalent width EW$>1$~keV. The continuum
is also significantly flatter than in the XMM~1 and XMM~3
5-10 keV spectra (at lower energies, the XMM~1 and XMM~3
spectra are inverted, due to a photoelectric cut off).
These features
strongly suggest that during the XMM~2 observation the source 
was in a reflection-dominated state.

One more interesting aspect is the constancy (within 5\%) of the soft emission.
This suggests that the soft spectrum is dominated by a component
not directly related to the primary AGN emission.
This is confirmed by the visual analysis of the {\em Chandra}
spectra of the central 2" region, and of the 30" region:
most of the soft emission comes from the extended region
between 2" and 30" (corresponding to a sphere of outer radius 
of $\sim2$~kpc).

In order to confirm the results of this visual inspection, we performed a detailed spectral 
analysis of each observation. The complete report on this work will be presented
elsewhere (Risaliti et al. 2005, in prep.). In the following we 
briefly summarize the results relevant for our interpretation of the 
continuum variations. 

\noindent{\bf {\em Chandra.}} The {\em Chandra} spectrum obtained from
the central 2" is unusually flat: the photon index of a simple power law model
is $\Gamma=-0.1\pm0.1$. A good fit is obtained with a cold reflection model ({\tt PEXRAV}
in XSPEC), plus an iron
emission line at 6.4~keV (equivalent width EW=200$_{-120}^{+150}$~eV). A
soft thermal component is not required by the data.  The spectrum
obtained from the larger 30" radius region can be fitted with the same
components, plus thermal emission with kT$\sim$0.8~keV, with a
normalization consistent with that found in all the {\em XMM-Newton}
observations. This implies that a) the hard emission comes from the
central region; b) the soft component originates from a larger nuclear
region (radius $\sim 1$~kpc).

\noindent{\bf XMM~1}. The first XMM spectrum, obtained only 3 weeks
after the {\em Chandra} observation, is clearly dominated by the
direct emission of the AGN above $\sim3$~keV. The best fit model
consists of all the components used to fit the {\em Chandra} spectrum,
plus an absorbed power law, completely dominating the emission
above a photoelectric cut off at $\sim3$~keV (
$N_H=4.8^{+0.3}_{-0.2}\times10^{23}$~cm$^{-2}$)
The best fit parameters for the cold reflection are fully consistent with those
found in the {\em Chandra} observation.
The ratio between the normalizations of the
reflection component and the power law is R=1.2. 

\noindent{\bf XMM 2}. The second XMM spectrum was obtained 3 weeks
after the first. The spectrum is now back in a reflection-dominated
state (Fig.~1). The best fit model consists of two components: a cold 
reflection emission, and an iron emission line at E=6.4~keV, with EW$\sim$1.2~keV.
All the parameters of the cold reflection component
are consistent with those found in
the previous XMM~1 and {\em Chandra} spectra.

\noindent{\bf XMM 3}. This spectrum, obtained 5 months after the XMM~2
observation, is dominated by the direct emission from the central
source. The best fit model is the same as in XMM~1, 
with a normalization of the direct component higher by 50\%,
and an absorbing column density
N$_H=3.4^{+0.2}_{-0.2}\times10^{23}$~cm$^{-2}$. 
The reflection component is consistent with that of XMM~1 and XMM~2.
The ratio between the normalizations of the
reflection component and the power law is R=0.8. 

The fit is significantly improved ($\Delta\chi^2=45$) by the addition
of a broad (possibly relativistic) iron emission line\footnote{This is 
an important point in itself, however it does not affect our treatment
of the continuum variations, therefore we will not discuss it further here, but
in Risaliti et al.~(2005).},
although other models, e.g. partial covering and a second, lower $N_H$, absorber
(as in NGC~4151, Zdziarski et al.~2002) may also fit.

\section{Discussion}

The spectral analysis of multiple {\em Chandra} and XMM-Newton
observations of NGC~1365 reveals
transitions from reflection-dominated to transmission-dominated,
on time scales no longer than 3 weeks.
This is, by a factor of 30, the shortest time scale observed so far for
such extreme X-ray spectral variations (Matt et al. 2003)
While we also find a kiloparsec scale
diffuse origin for the soft thermal component and the possible
presence of a relativistic iron line in the highest statistics
observation, we will defer discussing of these issues to where we
can also consider 
two new, longer ($\sim60$~ksec) {\em XMM-Newton} observations
(Risaliti et al. 2005, in prep.). Here we concentrate 
on the fast spectral transitions.

\noindent
The extreme column density variations observed from the {\em Chandra} and
XMM~1 observations, and from XMM~1 and XMM~2 can be due to either:
(A) fading of the central source, down to a flux 50 times or more
fainter than in the ``transmission dominated'' observations;
(B) increased ionization making the absorber more transparent;
(C) column density variability, with the reflection dominated spectra
having $N_H>10^{24}$~cm$^{-2}$ along the line of sight, 

{\bf (A) Fading of the central source.} In the intrinsic variability 
scenario, an almost complete switch-off of the source (down to a flux
$<2$\% of the intrinsic flux in the XMM~1 observation) is needed, in
a time shorter than 3 weeks. This can be compared with the typical cooling time
of the inner part of the accretion disk\footnote{Here we assume that
the fading in the X-rays is due to a decrease of the seed photons
from the disk which are up-scattered by the coronal hot electrons.}.
The thermal timescale in a Shakura-Sunyaev (1972) disk is approximately
$t_{th}\sim (\alpha\times\Omega_K)^{-1}$ 
(Frank, King \& Raine 2002, Starling et al. 2004), where $\alpha$ is the
viscosity parameter ($\alpha<0.1$ for a realistic model), and 
$\Omega_K$ is the Keplerian angular velocity, $\Omega_K=\sqrt{GM_{BH}/R^3}$.
The black hole mass, $M_{BH}$ can be estimated from the K magnitude 
of the galaxy (K=8.4, Jarrett et al. 2003), through the relation
between black hole mass and luminosity in the K band (Marconi \& Hunt 2003).
We obtain $M_{BH}\sim1.5\times10^8~M_\odot$, and 
 $t_{th} > 30~R_{20}^{3/2}$~days, where $R_{20}$ is the linear
dimension of the X-ray source in units of $20~R_S$.
Given the uncertain assumptions this result is in
marginal agreement with our observational requirement
$t<21$~days. However, this requires an  
extreme scenario, i.e. the complete switch off of the source right after
the XMM~1 observation. We conclude that intrinsic variability due to disk cooling 
is an unlikely explanation for the observed spectral changes.
However, in order to rule it out completely, shorter 
variations (in a timescale of a few days) should be observed.\\
{\bf (B) Ionization changes}. In order to make the cold absorber
found in the {\em Chandra} and XMM~2 observations transparent, an
increase of the ionization parameter $\xi$\footnote{
$\xi=L(2-10~keV)/nR^2$, where $n$ is the number density of the absorbing gas.} 
by at least $\sim3$ orders of magnitude
would be needed (from $\xi<1$, as required in order to completely 
obscure the 1-10~keV emission, to $\xi>10^3$, as required in order to make
the gas transparent. This would in turn imply a similar increase in
the intrinsic luminosity, which is not seen.\\ 
{\bf (C) $N_H$ variability}. 
We assume (as in Risaliti et al. 2002, Elvis et al. 2004) that the
absorber is made up of gas clouds moving around the central source
with Keplerian velocity (motions close to Keplerian dominate in the
high ionization parts of the BELR in AGNs, Peterson \& Wandel 2000).
The variability timescale is then given, to a first approximation, by
the crossing time for a cloud across the line of sight. 
Assuming
spherical clouds, the distance $R$ of the absorber
from the center is (Risaliti et al.~2002): 
$
R\sim600 t_{100}^2 n_{10}^2 N_{H,24}^{-2} R_S
$
where $t_{100}$ is the variability time in units of 100 ksec,
$n_{10}$ is the cloud density in units of 10$^{10}$~cm$^{-3}$
and $N_{H,24}$ is the column density of a single cloud in
units of 10$^{24}$~cm$^{-2}$.
Assuming that the variations from Compton thick to Compton thin
states were due to
a single cloud passing along the line of sight, the measured $N_H$
variation, $\Delta(N_H)=10^{24}$~cm$^{-2}$, corresponds to the
column density of a single cloud. 
 We further assume the cloud
crossing time to be no longer than the time interval between the XMM~1 and XMM~2
observations, $t_{100} \leq 20$.
With these numbers we obtain
$R \leq 2\times10^5 n_{10}^2 R_S$. 
A further stringent requirement for
the absorbing cloud is to be large enough to cover the X-ray source.
Using $M_{BH}=1.5\times10^8~M_{\odot}$ (see above), and assuming again a linear dimension
of the X-ray source D$\sim$20~$R_S$, we obtain $n<2\times10^9$~cm$^{-3}$ and
$R\leq10^4~R_S=4.5\times10^{17}$~cm. For comparison, the dust sublimation
radius (assuming the standard X-ray to bolometric correction of Elvis et al.~1994)
is of the order of $10^{17}$~cm.\\
The Compton thick - Compton thin variation discussed here is, by 
a factor $\sim$30,
the fastest observed so far (Matt et al. 2003, Guainazzi et al. 2004). 
Therefore, the arguments used to rule out
the intrinsic variability scenario are significantly less stringent in the other
known cases. As a consequence, it is at present impossible to tell
whether the observed extreme variability in NGC~1365 is an unique
case or is representative of the other known state-changing sources. 
However, several general considerations
can be made on the occurence of such variations, which are relevant
in modeling the circumnuclear medium of AGN, and in driving further
studies:\\
- Compton thick gas seems to be present in (almost) all X-ray obscured 
Seyfert Galaxies.
In $\sim$40-50\% of the sources it covers the line of sight to the central
X-ray source (Risaliti, Maiolino \& Salvati 1999). In the other X-ray obscured 
Seyfert Galaxies bright enough to allow a careful X-ray spectral analysis, the
high reflection efficiency suggests the presence of Compton thick gas
reflecting the primary emission (see Introduction).\\
- Even if we ascribe all the known thick-thin transitions to $N_H$ variability, 
this event is still quite rare: 
only four objects are known to show such variations (Matt et al.~2003) out
of the more than 30 known Compton thick AGNs (Comastri 2004). Recently
Guainazzi et al. (2004) found one more example (NGC~4939) in a sample of 11 objects.\\
- The few indications available on short time scale variations (for example
this work, and Elvis et al. 2004) suggest that the absorber is rather
compact, at a distance of $10^3-10^4~R_S$ from the central source.

This observational evidence can be explained if we assume a stratified structure 
for the circumnuclear absorber/reflector, with a central planar Compton-thick region 
and a decreasing average column density at greater angles or distance above the disk
(Fig.~2).
The gas is clumpy and at the distance of the order of that found in NGC~1365
($\sim10^4~R_S$). The average number of clouds along a given line
of sight is $N\sim5-10$ (except, possibly, for the completely
Compton thick region), in order to reproduce the observed average
$N_H$ variability in bright AGNs (Risaliti et al.~2002).
The angles have been chosen in order to reproduce the observed 
column density distribution of Risaliti et al.~(1999), with fractions
of 45\%, 25\% and 10\% of the solid angle covered by gas with 
column density $N_H>10^{24}$~cm$^{-2}$, $N_H\sim10^{23}-10^{24}$~cm$^{-2}$,
and $N_H\sim10^{22}-10^{23}$~cm$^{-2}$, respectively, and with the remaining
20\% free from absorption, in agreement with the obscured/unobscured Seyferts
ratio of Maiolino \& Rieke (1995).

In this scheme strong absorption variations are unlikely in the extreme
cases of heavily Compton-thick sources (line of sight through the central region
of the absorber) and low obscuration ($N_H=10^{22}-10^{23}$~cm$^{-2}$) sources 
(line of sight far from the absorber),
while they can happen if the line of sight intersects the transition region between the
Compton thick and Compton thin zones. In this simple scheme, NGC~1365 is
one of the sources seen through this ``transition region''.
We note that a similar stratified and compact structure has been proposed 
to explain the relative X-ray/mid IR absorption properties of Luminous
Infrared Galaxies (Risaliti et al.~2000), and to explain the $N_H$
distribution and the megamaser emission
in AGNs (Kartje, K{\"o}nigl \& Elitzur 1999). In this latter case the distribution
arises as the result of a disk wind.

The proposed scheme needs to be tested with more detailed studies of the
single sources showing the most extreme variability (a new 
{\em XMM-Newton} observational campaign is on going) and with an unbiased
statistical analysis of the occurence of such extreme variations.

\acknowledgements

We are grateful to A. Siemiginowska and G. Matt for useful discussions.
This work was partially supported by NASA grants NAG5-13161,
NNG04GF97G, and NAG5-16932.


\begin{figure}
\epsscale{0.8}
\plotone{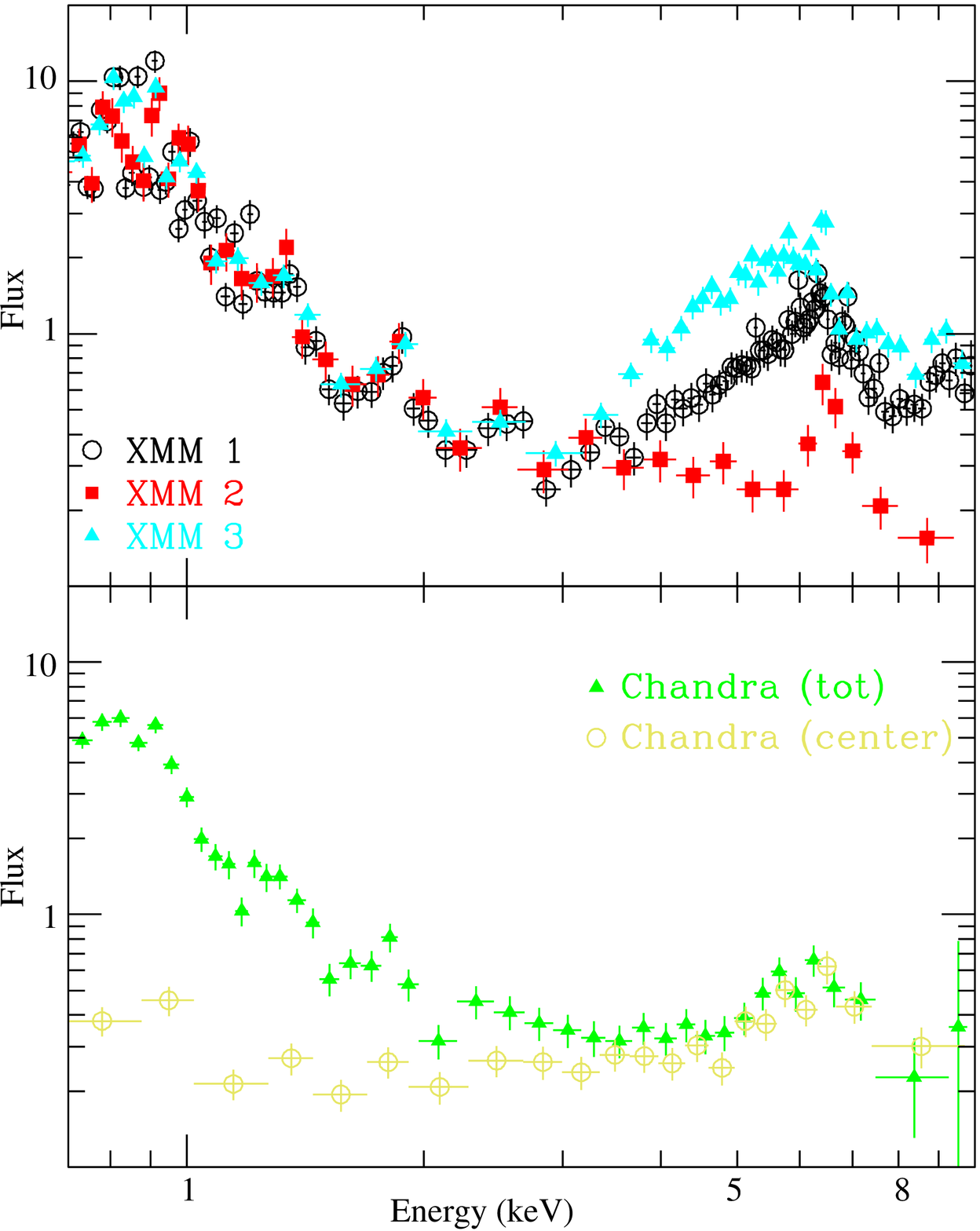} 
\caption{Upper panel: Unfolded spectra (using the best
fit models of Tab.~1) of the XMM~1, XMM~2 and XMM~3
observations. Central Panel: unfolded spectra of the Chandra observation, extracted
from a circular region of 2 arcsec (circles) and of 30 arcsec (triangles),
as for the XMM observations. 
Fluxes are in units of $10^{-4}$~erg~s$^{-1}$~cm$^{-2}$~keV$^{-1}$.}
\end{figure}

\begin{figure}
\epsscale{0.8} 
\plotone{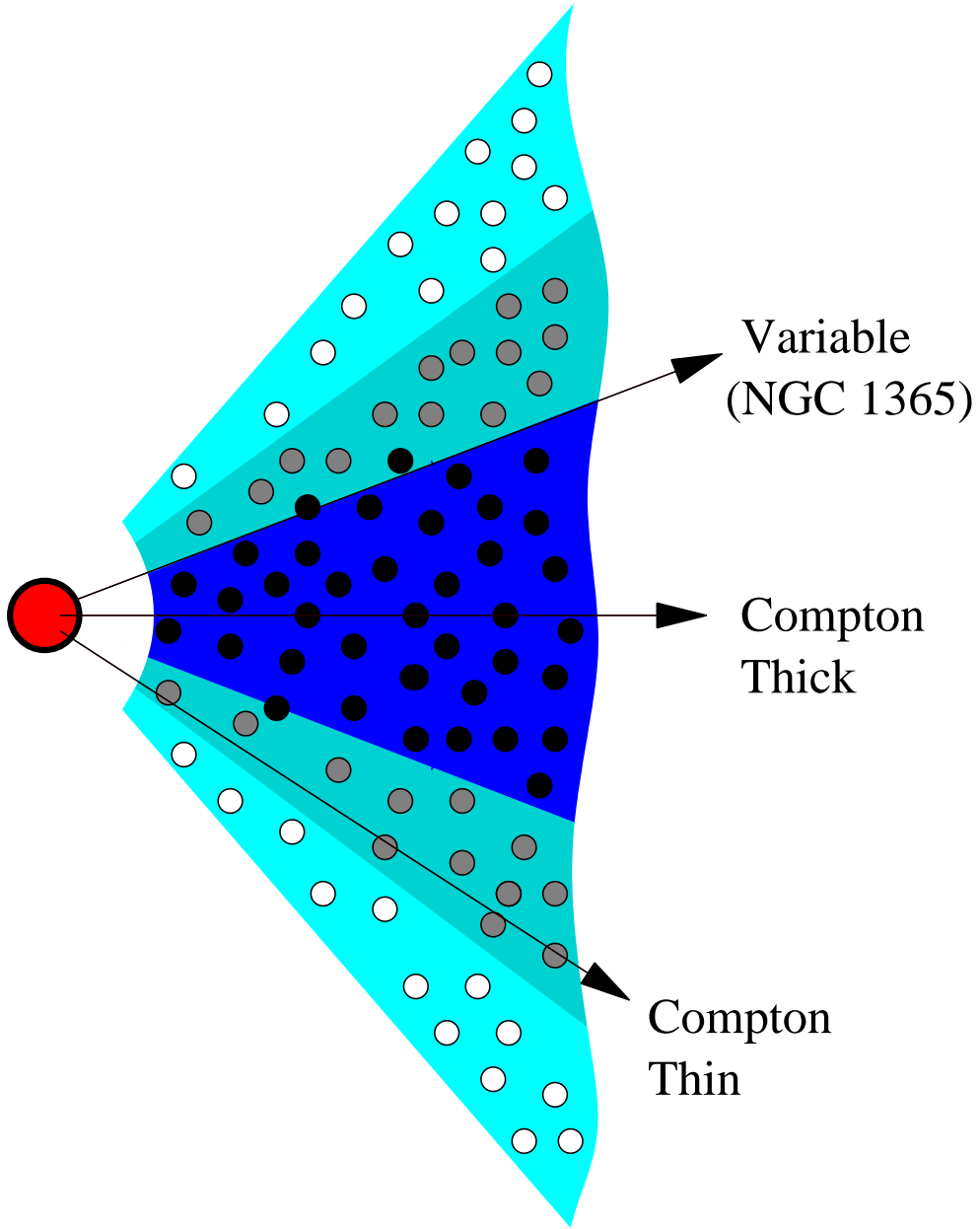}
\caption{Section of the proposed structure of the circumnuclear
absorber/reflector.
The average column density is indicated by the 
colors of the background and of the clouds (dark: $N_H>10^{24}$~cm$^{-2}$;
medium: $10^{23}$~cm$^{-2}<N_H<10^{24}$~cm$^{-2}$; light: 
$10^{22}$~cm$^{-2}<N_H<10^{23}$~cm$^{-2}$). The angles are chosen in order
to reproduce the column density distribution of Risaliti et al. (1999).}
\end{figure}

\clearpage

\begin{table*}
\caption{NGC 1365 - X-ray Observation log. \& Spectral Fitting Results}
\centerline{\begin{tabular}{lccccccccc}
\hline
Observatory & Date & Counts & T$^a$ & $\Gamma$ & N$_H^b$ & $R^c$ & F(2-10)$^c$ &L(2-10)$^d$ & $\chi^2_r$\\
\hline
{\em Chandra}$^f$ & 2002 Dec 24 &7478 & 14.6 & 1.98$^{+0.13}_{-0.14}$(g) & --                  & --  & 0.21 & 0.14 & 1.09 \\
XMM-1 & 2003 Jan 16             &14866& 17.8 & 2.06$^{+0.06}_{-0.03}$    &47.6$^{+3.0}_{-1.5}$ & 1.2 & 0.48 & 1.23 & 1.09 \\
XMM-2 & 2003 Feb 09             &3503 & 5.8  & 2.09$^{+0.11}_{-0.14}$(g) & --                  & --  & 0.17 & 0.11 & 1.06\\
XMM-3 & 2003 Aug 13             &9466 & 8.7  & 2.33$^{+0.12}_{-0.12}$    &33.5$^{+1.9}_{-2.0}$ & 0.8 & 0.70 & 2.30 & 1.21 \\
\hline
\end{tabular}}

$^a$: Duration (ksec).
$^b$: in units of 10$^{22}$~cm$^{-2}$. 
$^c$: Ratio between the normalizations of the reflected and transmitted components.
$^d$: Observed 2-10~keV flux, in units of 10$^{-11}$~erg~s$^{-1}$~cm$^{-2}$. 
$^e$: 2-10~keV luminosity in units of $10^{42}$~erg~s$^{-1}$. 
For the transmission dominated spectra, the luminosity is absorption corrected.
%
$^f$: 2 arcsec dia. aperture.
$^g$: photon index of the intrinsic power law in the PEXRAV model.
\end{table*}


\begin{thebibliography}{}
\bibitem[Comastri]{} Comastri, A. 2004, in ``Supermassive Black Holes in the Distant
Universe'', ed. A. J. Barger, (Boston: Kluwer Academic Publishers),
308, 245-172
\bibitem[Elvis et al.(2004)]{587} Elvis, M., Risaliti, G., Nicastro, F.,
Miller, J., \& Puccetti, S. 2004, \apj, 635, L25
\bibitem[Frank, King, \& Raine(2002)]{2002apa..book.....F} Frank, J., King, 
A., \& Raine, D.~J.\ 2002, Accretion Power in Astrophysics: Third Edition, 
by Juhan Frank, Andrew King, and Derek J.~Raine.~Cambridge University 
Press, 2002
\bibitem[Ghisellini, Haardt, \& Matt(1994)]{1994MNRAS.267..743G} 
Ghisellini, G., Haardt, F., \& Matt, G.\ 1994, \mnras, 267, 743 
\bibitem[Gilli et al.(2000)]{2000A&A...355..485G} Gilli, R., Maiolino, R., 
Marconi, A., Risaliti, G., Dadina, M., Weaver, K.~A., \& Colbert, E.~J.~M.\ 
2000, \aap, 355, 485
\bibitem[Guainazzi(2002)]{2002MNRAS.329L..13G} Guainazzi, M.\ 2002, \mnras, 
329, L13
\bibitem[Guainazzi et al.(2004)]{} Guainazzi, M., Fabian, A. ~C., Iwasawa, K., 
Matt, G., \& Fiore, F. 2004, \mnras, in press (astro-ph/0409689) 
\bibitem[Iyomoto et al.(1997)]{1997PASJ...49..425I} Iyomoto, N., Makishima, 
K., Fukazawa, Y., Tashiro, M., \& Ishisaki, Y.\ 1997, \pasj, 49, 425 
\bibitem[Jarrett et al.(2003)]{2003AJ....125..525J} Jarrett, T.~H., 
Chester, T., Cutri, R., Schneider, S.~E., \& Huchra, J.~P.\ 2003, \aj, 125, 
525 
\bibitem[Kartje]{} Kartje, J. ~P., K{\" o}nigl, A., \& Elitzur, M. 1999, 
\apj, 513, 180
\bibitem[Krolik \& Begelman(1988)]{1988ApJ...329..702K} Krolik, J.~H.~\&
Begelman, M.~C.\ 1988, \apj, 329, 702
\bibitem[Magdziarz \& Zdziarski(1995)]{1995MNRAS.273..837M} Magdziarz, 
P.~\& Zdziarski, A.~A.\ 1995, \mnras, 273, 837 
\bibitem[Marconi \& Hunt(2003)]{2003ApJ...589L..21M} Marconi, A.~\& Hunt, 
L.~K.\ 2003, \apjl, 589, L21 
\bibitem[Matt, Guainazzi, \& Maiolino(2003)]{2003MNRAS.342..422M} Matt, G., 
Guainazzi, M., \& Maiolino, R.\ 2003, \mnras, 342, 422 
\bibitem[Peterson \& Wandel(2000)]{2000ApJ...540L..13P} Peterson, B.~M.~\& 
Wandel, A.\ 2000, \apjl, 540, L13 
\bibitem[]{610} Puccetti, S., Risaliti, G., Fiore, F., Elvis, M., Nicastro, F.,
Perola, G.C., Capalbi, M. 2003, Proc. of the BeppoSAX Symposium, The
Restless High-Energy Universe, E.P.J. van den Heuvel, J.J.M. in 't
Zand,  and R.A.M.J. Wijers (Eds), astro-ph/0311446
\bibitem[Risaliti et al.(1999)]{1999ApJ...522..157R} Risaliti, G., 
Maiolino, R., \& Salvati, M.\ 1999, \apj, 522, 157 
\bibitem[Risaliti, Maiolino, \& Bassani(2000)]{2000A&A...356...33R} 
Risaliti, G., Maiolino, R., \& Bassani, L.\ 2000, \aap, 356, 33 
\bibitem[Risaliti(2002)]{2002A&A...386..379R} Risaliti, G.\ 2002, \aap, 
386, 379 
\bibitem[Risaliti, Elvis, \& Nicastro(2002)]{2002ApJ...571..234R} Risaliti, 
G., Elvis, M., \& Nicastro, F.\ 2002, \apj, 571, 234
\bibitem[Shakura \& Sunyaev(1973)]{1973A&A....24..337S} Shakura, N.~I., \& 
Sunyaev, R.~A.\ 1973, \aap, 24, 337 
\bibitem[Spergel et al.(2003)]{2003ApJS..148..175S} Spergel, D.~N., et al.\ 
2003, \apjs, 148, 175 
\bibitem[Starling, Siemiginowska, Uttley, \& 
Soria(2004)]{2004MNRAS.347...67S} Starling, R.~L.~C., Siemiginowska, A., 
Uttley, P., \& Soria, R.\ 2004, \mnras, 347, 67 
\bibitem[Urry \& Padovani(1995)]{1995PASP..107..803U} Urry, C.~M.~\&
Padovani, P.\ 1995, \pasp, 107, 803
\bibitem[Zdziarski et al.(2002)]{2002ApJ...573..505Z} Zdziarski, A.~A., 
Leighly, K.~M., Matsuoka, M., Cappi, M., \& Mihara, T.\ 2002, \apj, 573, 
505 
\end{thebibliography}
\end{document}